\documentclass[]{spie}  

\usepackage{amsmath,amsfonts,amssymb}
\usepackage{graphicx}
\usepackage[colorlinks=true, allcolors=blue]{hyperref}

\title{New concepts in vector-Apodizing Phase Plate coronagraphy}

\author[a]{Steven P. Bos}
\author[a]{David S. Doelman}
\author[a]{Kelsey L. Miller}
\author[a]{Frans Snik}
\affil[a]{Leiden Observatory, Leiden University, P.O. Box 9513, 2300 RA Leiden, The Netherlands}

\authorinfo{email: stevenbos@strw.leidenuniv.nl}

\begin{document} 
\maketitle

\begin{abstract}
The vector-Apodizing Phase Plate (vAPP) is a pupil-plane coronagraph that manipulates phase to create dark-holes in the stellar PSF. 
The phase is induced on the circular polarization states through the inherently achromatic geometric phase by spatially varying the fast axis orientation of a half-wave liquid-crystal layer. 
The two polarized PSFs can be separated, either by a quarter-wave plate (QWP) followed by a polarizing beamsplitter (PBS) for broadband operation, or a polarization sensitive grating (PSG) for narrowband or IFS operation.
Here we present new vAPP concepts that lift the restrictions of previous designs and report on their performance. 
We demonstrated that the QWP+PBS combination puts tight tolerances on the components to prevent leakage of non-coronagraphic light into the dark-hole. 
We present a new broadband design using an innovative two-stage patterned liquid-crystal element system based on multi-color holography, alleviating the leakage problem and relaxing manufacturing tolerances. 
Furthermore, we have shown that focal-plane wavefront sensing (FPWFS) can be integrated into the vAPP by an asymmetric pupil. 
However, such vAPPs suffer from a reduced throughput and have only been demonstrated with a PSG in narrowband operation. 
We present advanced designs that maintain throughput and enable phase and amplitude wavefront sensing. 
We also present broadband vAPP FPWFS designs and outline a broadband FPWFS algorithm. 
Finally, previous dual-beam vAPP designs for sensitive polarimetry with one-sided dark holes were very complex. 
We show new dual-beam designs that significantly reduce the complexity. 
\end{abstract}


\section{Introduction}\label{sec:intro}
The exploration of circumstellar environments to search for exoplanets and signs of their formation is an exciting and rapidly expanding research field. 
However, it is also a field that needs to overcome many technical challenges before an Earth-like exoplanet can be imaged and characterized to look for signs of life. 
This is because exoplanets are much fainter than their host stars, and are located at small angular separations. 
For example, if an Earth-like exoplanet would be imaged around a Sun-like star from a distance of 10 pc in the visible wavelength ranges ($\sim$0.3-1 $\mu$m), it would be at a contrast of $\sim$$10^{-10}$, and angular separation of $\sim$100 mas\cite{traub2010direct} . 
Current ground-based high-contrast imaging (HCI) instruments \cite{macintosh2014first, jovanovic2015subaru, males2018magao, beuzit2019sphere} consist of extreme adaptive optics systems to measure and correct wavefront aberrations caused by the Earth's atmosphere, advanced coronagraphs to suppress starlight\cite{guyon2006theoretical}, and use various observing techniques\cite{kuhn2001imaging, sparks2002imaging, marois2006angular, ruane2019reference} to remove speckle noise in post processing. 
A host of different focal-plane wavefront sensors\cite{jovanovic2018review} (FPWFS) are being developed to measure and correct non-common path aberrations between the science focal plane and the main wavefront sensor, but are generally not yet included in science observations. 
The submodules of most current instruments are designed and optimized independently, while ideally the entire HCI system is optimized as whole, integrating all submodules, to get the best possible performance. 
For example, in VLT/SPHERE\cite{beuzit2019sphere} NCPA measurements with ZELDA\cite{vigan2018sky} cannot be done simultaneously with coronagraphic observations, lowering the science duty cycle of the instrument.  
It is much more efficient if the FPWFS and coronagraph would be combined into one device. \\
\begin{figure}[!htb]
\begin{center}
\begin{tabular}{c}
\includegraphics[width=0.75\textwidth]{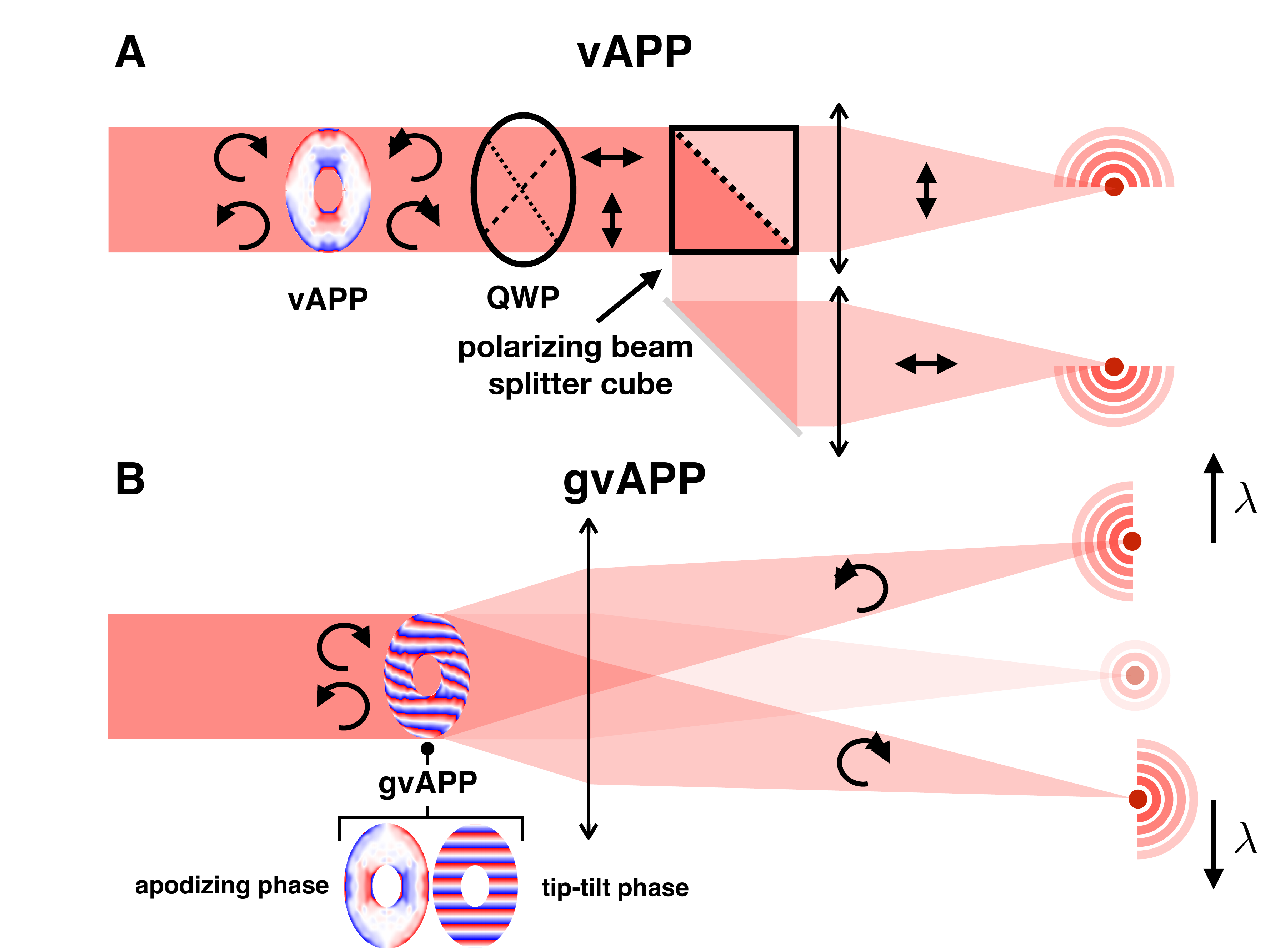}
\end{tabular}
\end{center}
\caption{
Schematics of a) the vector-Apodizing Phase Plate (vAPP) and b) the grating-vAPP (gvAPP). 
Adopted from Ref~\citenum{bos2018fully}.
}
\label{fig:ExplanationvAPPgvAPP}  
\end{figure} 

In this work we present new vector-Apodizing Phase Plate (vAPP) designs that combine coronagraphy with broadband imaging, focal-plane wavefront sensing, polarimetry, and interferometry.  
The vAPP\cite{snik2012vector, otten2014vector, otten2014performance, otten2017sky} is a coronagraph that manipulates the pupil-plane phase to cancel light in a selected region of the stellar point spread function (PSF). 
We refer to these regions as dark holes. 
The vAPP introduces phase to the circular polarization states by the achromatic geometric phase \cite{pancharatnam1956generalized, berry1987adiabatic}. 
The geometric phase is induced by a patterned half-wave liquid-crystal layer that has a carefully designed\cite{por2017optimal} spatially varying fast-axis angle\cite{miskiewicz2014direct}. 
While the geometric phase is inherently achromatic due to geometric effects, the efficiency with which the phase is acquired by the light depends on the retardance of the liquid-crystal layer, which can be chromatic. 
The light that does not acquire the desired phase will form a non-coronagraphic PSF,  and this is referred to as leakage. 
If the leakage becomes too strong, it can have detrimental effects on the contrast in the dark hole.  
By combining multiple layers of self-aligning liquid crystals, devices can be designed that have low leakage over broad wavelength ranges\cite{komanduri2012multi, komanduri2013multi}.   
Due to the nature of the geometric phase, both circular polarization states receive an equal but opposite phase, which results into two coronagraphic PSFs with opposite dark holes.  
These can be separated by a combination of a quarter-wave plate (QWP) and a polarizing beamsplitter, as shown in \autoref{fig:ExplanationvAPPgvAPP}a.
The simplest and most common implementation is the grating-vAPP\cite{otten2014vector} (gvAPP), which integrates a polarization-sensitive grating\cite{oh2008achromatic} into the vAPP design.
This is shown in \autoref{fig:ExplanationvAPPgvAPP}b.
To prevent smearing by the grating, gvAPPs are mainly used with narrowband filters or integral-field spectrographs. \\

Here we give a brief overview of previous efforts to integrate the vAPP with other submodules of HCI instruments. 
By adding a polarization modulator and QWP upstream of the vAPP, it becomes sensitive to all linear polarization states, enabling polarimetry\cite{snik2014combining}.  
When a micro-lens array, feeding single-mode fibers, is placed in the focal plane, the vAPP can be designed such that starlight is optimally rejected by the fibers, while the planetlight is still injected\cite{por2018single, haffert2018single}.  
These fibers can then feed high-resolution spectrographs, integrating the vAPP with spectroscopy. 
There have been multiple efforts to integrate the vAPP with focal-plane wavefront sensing. 
The coronagraphic modal wavefront sensor (cMWFS)\cite{wilby2017coronagraphic, por2016focal} encodes wavefront information by adding specifically designed holograms to the vAPP. 
More recently, we started to engineer the coronagraphic PSF such that it allows for wavefront sensing by adding pupil-plane amplitude asymmetries to the vAPP design\cite{bos2019focal}. 
We refer to these designs as Asymmetric Pupil vAPPs (APvAPP). 
This is a more efficient solution compared to the cMWFS when the aim is to control a large number of modes. \\

The outline of this work is as follows. 
In \autoref{sec:broadband_polarimetric_APvAPP} we present a broadband, polarimetric APvAPP design.
This design integrates broadband coronagraphy with polarimetry and wavefront sensing, but has very tight tolerances on the optical components, making it hard to manufacture and align.  
To alleviate the manufacturing problem, we present the multi-color vAPP in \autoref{sec:multi-color_vAPP}, which is a novel vAPP design that completely eliminates leakage problems.
For sensitive polarimetry, dual-beam polarimetric systems are required, but are very complex for vAPP systems\cite{snik2014combining}. 
To simplify dual-beam polarimetry with the vAPP, we sacrifice bandwidth and use a gvAPP to generate two beams, which is presented in \autoref{sec:dual-beam_gvAPP}.
Another solution is presented in \autoref{sec:unpol_gvAPP} with the unpolarized gAPvAPP, which has a design that makes the coronagraphic unpolarized, opposite to regular gvAPPs that have circular polarized coronagraphic PSFs. 
Current gvAPPs can only measure phase aberrations, but there are also amplitude aberrations, which are currently out of reach. 
In \autoref{sec:multiplexed_APvAPP} we present an APvAPP design that can measure both phase and amplitude. 
Then in \autoref{sec:gSAMvAPP}, we present the gSAMvAPP, a fusion of an APvAPP with a sparse aperture mask (SAM). 
Finally, we conclude in \autoref{sec:conclusion}. 
%
%
%
%
%
%
%
%
\section{Broadband polarimetric APvAPP}\label{sec:broadband_polarimetric_APvAPP}
\begin{figure}[!htb]
\begin{center}
\begin{tabular}{c}
\includegraphics[width=0.75\textwidth]{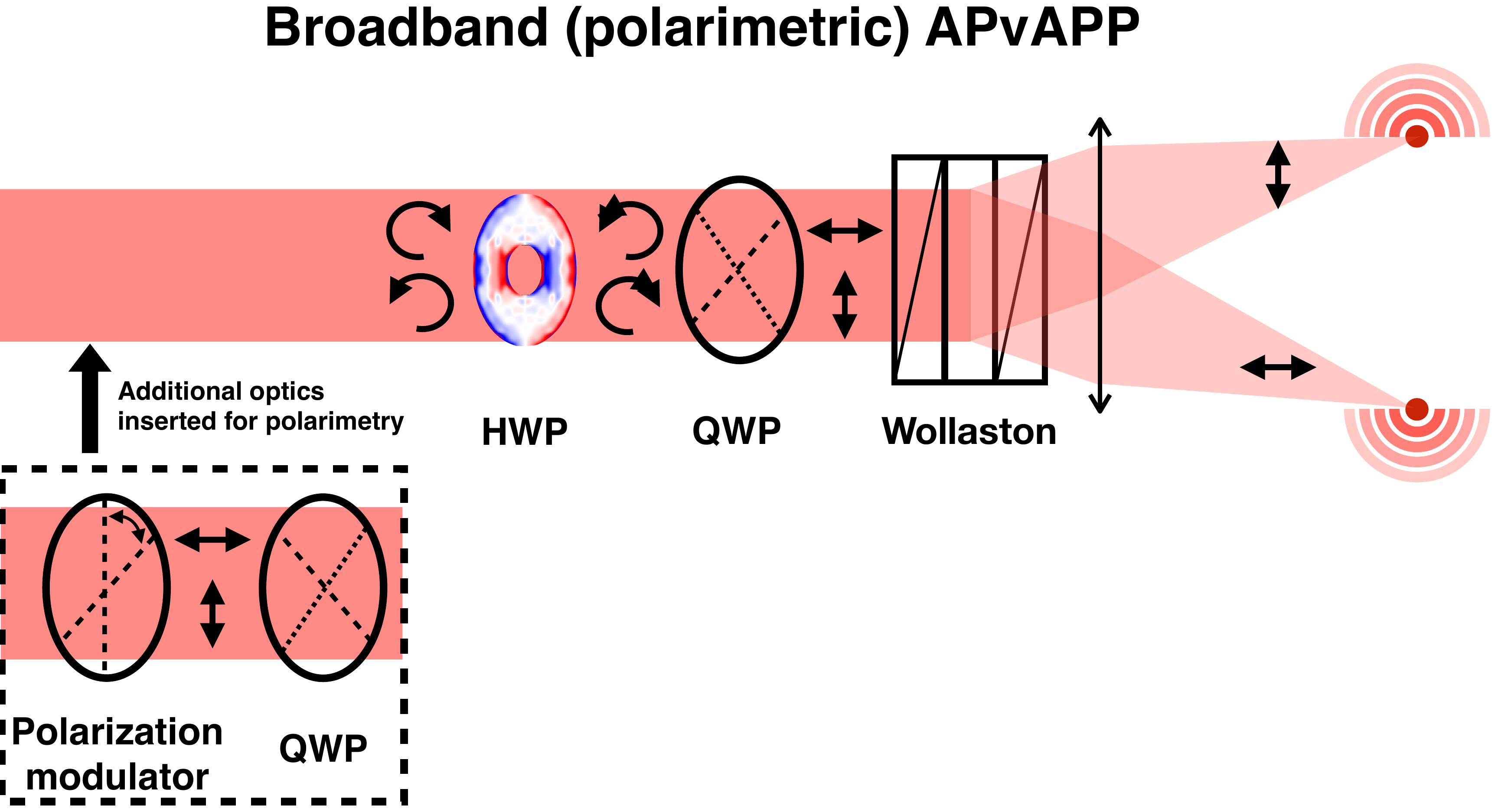}
\end{tabular}
\end{center}
\caption{
Schematic of a broadband (polarimetric) APvAPP. 
A polarization modulator (rotating HWP) and QWP are inserted to enable the polarimetric mode of the coronagraph. 
}
\label{fig:APvAPPschematic}  
\end{figure}

\begin{figure}[!htb]
\begin{center}
\begin{tabular}{c}
\includegraphics[width=0.75\textwidth]{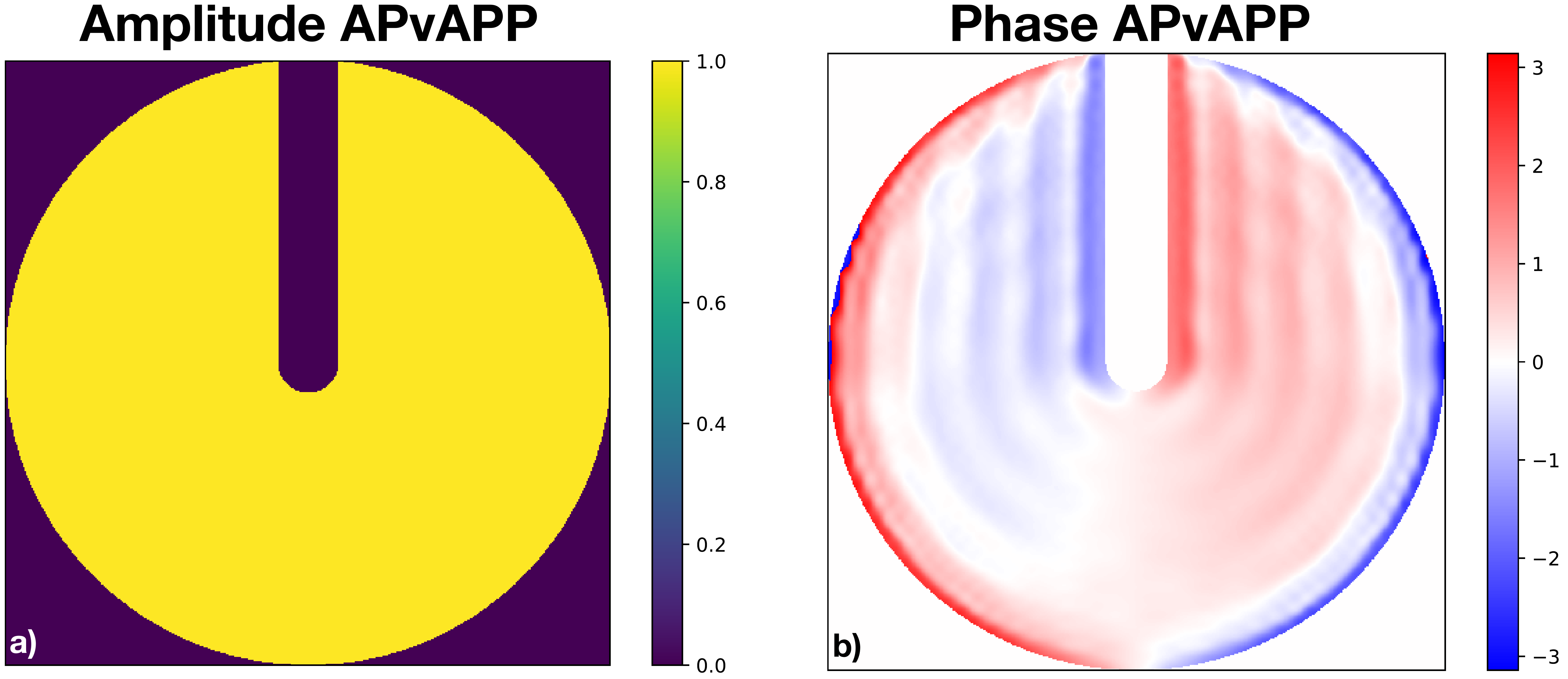}
\end{tabular}
\end{center}
\caption{Pupil-plane a) amplitude and b) phase design of the APvAPP. 
The amplitude shows a clear asymmetry, which enables wavefront sensing with the APvAPP.
}
\label{fig:APvAPPdesign}  
\end{figure}

\begin{figure}[!htb]
\begin{center}
\begin{tabular}{c}
\includegraphics[width=0.75\textwidth]{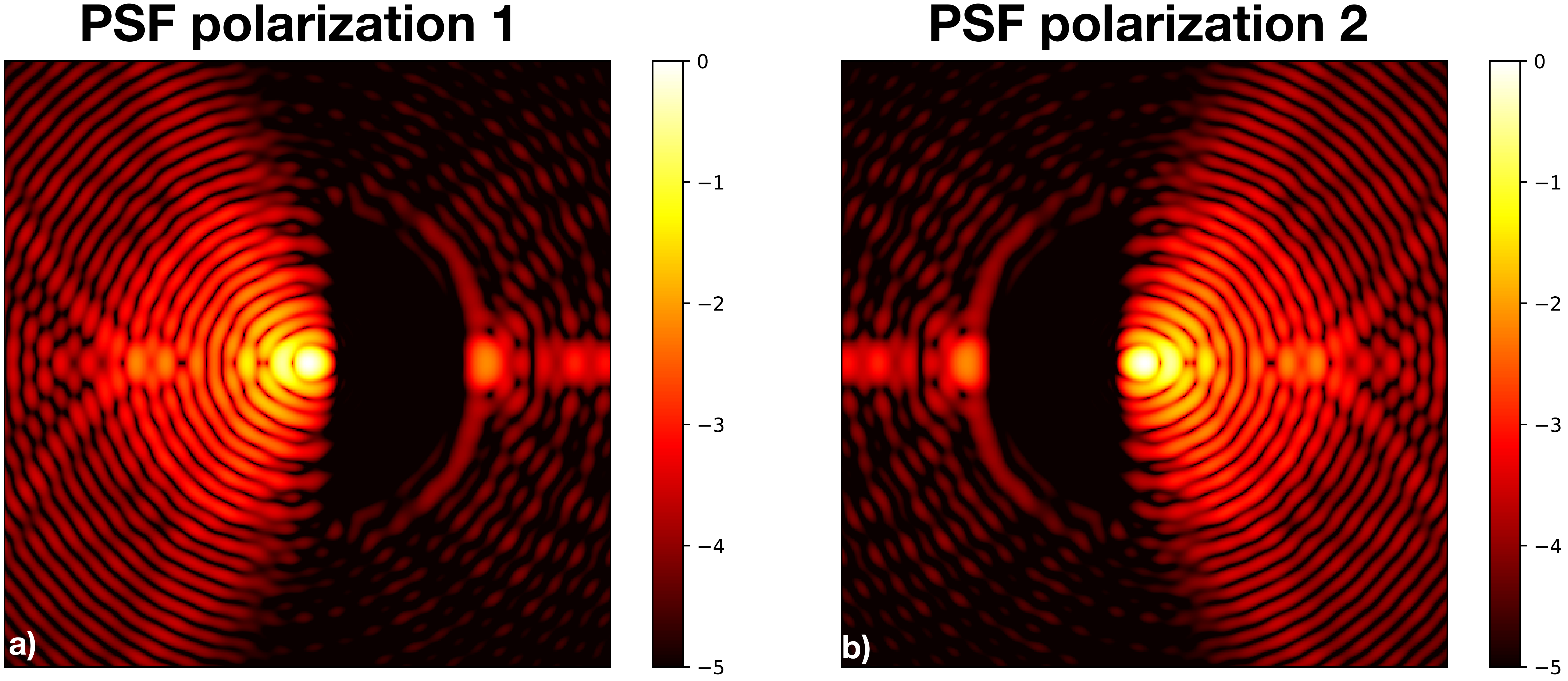}
\end{tabular}
\end{center}
\caption{The two coronagraphic PSFs for two orthogonal polarization states.
}
\label{fig:APvAPP_PSFs}  
\end{figure}

%
In \autoref{fig:APvAPPschematic} a schematic of the broadband polarimetric APvAPP is shown. 
Light comes in from the left and, when the polarimetric mode is turned on, first passes a polarization modulator, which modulates the polarization states measured by the vAPP. 
The linear polarization states are converted to circular polarization, which is necessary as the geometric phase operates on the circular polarization states. 
Then comes the liquid-crystal optic to apply the phase that generates the coronagraphic PSFs. 
The circular polarization is then converted back to linear by an additional QWP before a polarizing beamsplitter separates the coronagraphic PSFs. 
The phase and amplitude of the APvAPP are shown in \autoref{fig:APvAPPdesign}. 
The pupil-plane asymmetry that enables is wavefront sensing is clearly visible in \autoref{fig:APvAPPdesign}a.
The resulting coronagraphic PSFs for the two polarization states are shown in \autoref{fig:APvAPP_PSFs}. 
As there is no grating is involved, this design can in principle operate over broad wavelength ranges without suffering from wavelength smearing. 
The design as presented in \autoref{fig:APvAPPschematic} and \autoref{fig:APvAPPdesign} combines coronagraphy with broadband imaging, polarimetry and wavefront sensing. \\

A model-based wavefront sensing algorithm is presented in Ref~\citenum{bos2019focal}.
It estimates the wavefront by fitting a model of the APvAPP to an image acquired by the instrument. 
It minimizes the following objective function:
\begin{equation}\label{eq:objective_function}
\mathcal{L} (\alpha | {D}) = \sum_{{x}} \frac{1}{2 {\sigma}_n(x)^2} ({D(x)} - {M}(x, \alpha))^2 + \mathcal{R}(\alpha),
\end{equation}
with $D$ the image, $M$ the model of the APvAPP, $\sigma_n$ the noise, $\mathcal{R}(\alpha)$ a regularization term, and $\alpha$ a vector containing the amplitudes of the wavefront modes that are fitted to the model.  
The sum over $x$ is over all spatial pixels. 
The algorithm currently operates with narrowband images of gAPvAPPs. 
Which means that a monochromatic model as approximation of the gAPvAPP is sufficient. 
However, with a APvAPP operating over broad wavelengths ranges, this assumption will not hold. 
Therefore, the objective function presented in \autoref{eq:objective_function} needs to be adjusted to: 
\begin{equation}\label{eq:objective_function_broadband}
\mathcal{L} (\alpha | {D}) = \sum_{{x}} \frac{1}{2 {\sigma}_n(x)^2} \left({D(x)} - \sum_{\lambda} {M}(x, \alpha, \lambda) \right)^2 + \mathcal{R}(\alpha),
\end{equation}
with $\lambda$ the wavelength. 
The sum over $\lambda$ basically results in a broadband model of the APvAPP at the cost of more computational effort. 
For non-coronagraphic systems this method was successfully tested by Ref~\citenum{seldin2000closed}.
It also requires a more complex model of the APvAPP, including the QWPs, HWP and polarizing beamsplitter, and other relevant polarization effects of the system (e.g. instrumental polarization). \\

In previous work \cite{bos2018fully} we investigated the requirements on the individual components of a broadband polarimetric vAPP to limit the leakage to $10^{-5}$ in the dark hole. 
We found that the retardance offset of the liquid-crystal layer should be less than $3.6^\circ$, and that the retardance and fast-axis angle offset of the QWPs should be less than $0.8^\circ$ and $0.4^\circ$, respectively. 
These are extremely tight requirements, especially over broad wavelengths ranges such as an astronomical band. 
The leakage by the liquid-crystal optic can be minimized by using the so-called "double-grating" technique\cite{doelman2017patterned, doelman2020minimizing}. 
To reach the requirements for the QWPs, they need to be constructed from multilayer stacks of birefringent materials, each with tight alignment tolerances with respect to each other, complicating the manufacturing process. 
This will also make it harder to meet the wavefront error requirements as the QWPs will become thicker optics.
In the next section we present a vAPP design that completely eliminates these requirements by sacrificing other components of the design. 
%
%
\section{Multi-color vAPP}\label{sec:multi-color_vAPP}
\begin{figure}[!htb]
\begin{center}
\begin{tabular}{c}
\includegraphics[width=1\textwidth]{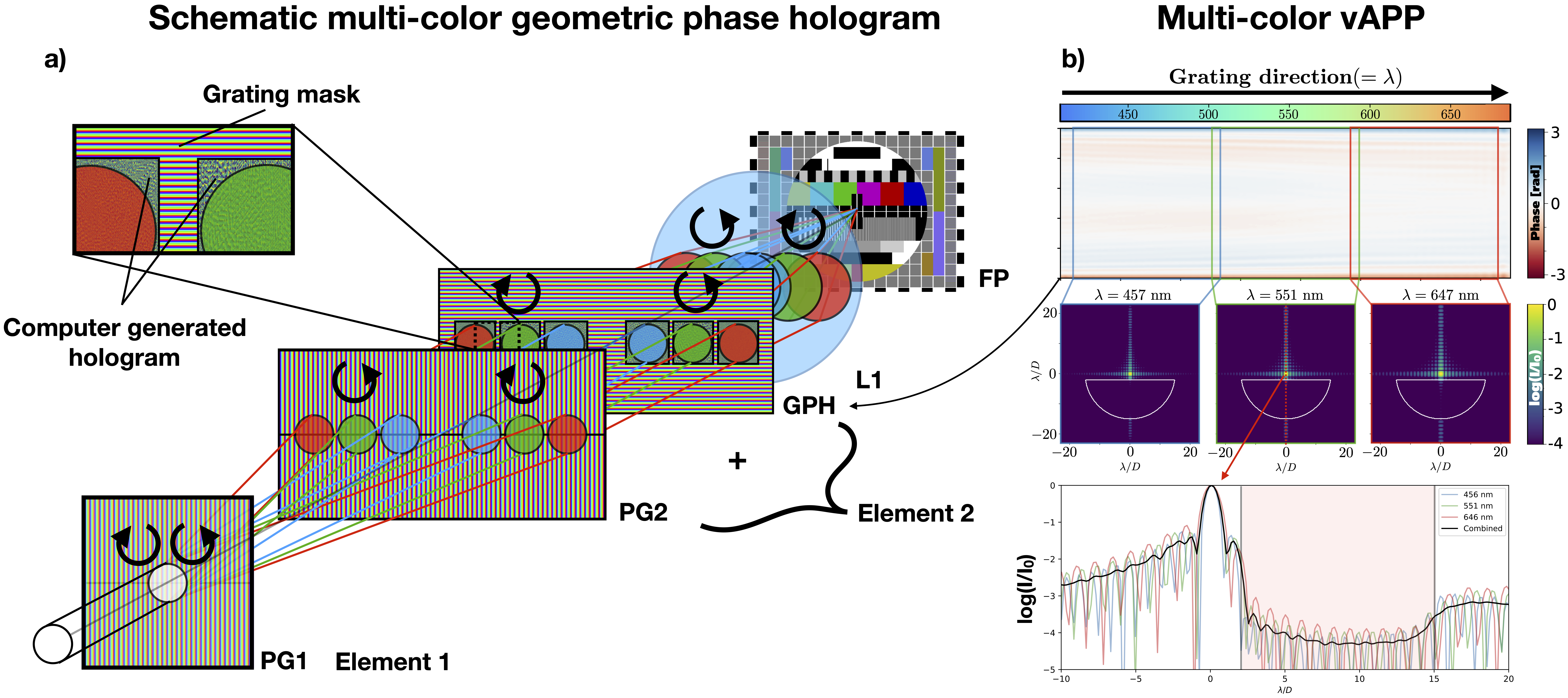}
\end{tabular}
\end{center}
\caption{a) Schematic of the multi-color geometric phase hologram principle. 
Figure adopted from Ref~\citenum{doelman2019multi}. 
b) The design of a multi-color vAPP in the upper panel, the resulting PSFs for three wavelengths in the middle panels, and the broadband PSF in the lower panel. 
}
\label{fig:WhiteLightvAPP}  
\end{figure}

%
As discussed in Ref~\citenum{bos2018fully} and the previous section, tight requirements on retardance and fast-axis angle offsets of QWPs and liquid-crystal plates prevents easy manufacturing of broadband vAPPs.  
In this section we present a novel broadband vAPP design that completely eliminates leakage problems by combining multiple liquid-crystal plates. 
This idea is based on the concept of multi-color geometric phase holograms\cite{doelman2019multi} (MCGPH). 
A schematic of a MCGPH is shown in \autoref{fig:WhiteLightvAPP}a. 
It consists of two stages: first a polarization grating\cite{oh2008achromatic} (PG) separates the light into the opposite circular polarization states. 
The PG is a diffractive element and will disperse the two beams into their constituent colors. 
When the beams hit the second stage, the colors are spatially separated and are directed through different holographic patterns. 
These holographic patterns collimate the light by a second PG, and simultaneously apply a phase pattern of choice. 
For example, the holograms can be designed such that every color generate an image of the same spatial size, which becomes a white light hologram for RGB input. \\ 

We refer to the vAPP equivalent of the MCGPH as the multi-color vAPP (MCvAPP). 
The MCvAPP follows the same recipe as described for the MCGPH, but will have a more complicated hologram design. 
The upper panel of \autoref{fig:WhiteLightvAPP}b shows the phase design of the MCvAPP for one circular polarization state.  
Along the horizontal direction (over which the colors are spatially separated) the design changes.  
This allows for a dark hole size that is constant (has a constant inner working angle in milliarcseconds) with wavelength. 
The other circular polarization state could have the same or a different phase design, whatever is required for the instrument. 
The PSFs generated for three different wavelengths are plotted in the middle panel of \autoref{fig:WhiteLightvAPP}b. 
It clearly shows that the dark holes for these wavelengths have the same spatial size. 
A radial cut of the coronagraphic PSF is shown in the lower panel of \autoref{fig:WhiteLightvAPP}b.
The two liquid-crystal plates making up the two stages will each create a leakage term. 
The leakage from the first stage will be on axis, while the two circular polarization states are diffracted off axis. 
This leakage can be easily removed by an opaque structure at the center of the second stage. 
As the light incident on the second stage arrives at an angle, the leakage from the second stage will also have this angle, making it easy to filter out by additional opaque structures downstream of the liquid-crystal plates. \\   

The MCvAPP is a new concept for broadband coronagraphy with the vAPP. 
The advantages are: a fixed dark hole size with wavelength (no spectral smearing), independent designs for the two circular polarization states, and elimination of leakage terms, strongly reducing the requirements on the optics. 
However, this comes at the cost of increased system complexity in other ways because the beam will be wider for part of the system, and the designs currently only work for rectangular apertures.  
We have manufactured a prototype that soon will be tested in the lab. 
\section{Dual-beam polarimetric gvAPP}\label{sec:dual-beam_gvAPP}
\begin{figure}[!htb]
\begin{center}
\begin{tabular}{c}
\includegraphics[width=0.75\textwidth]{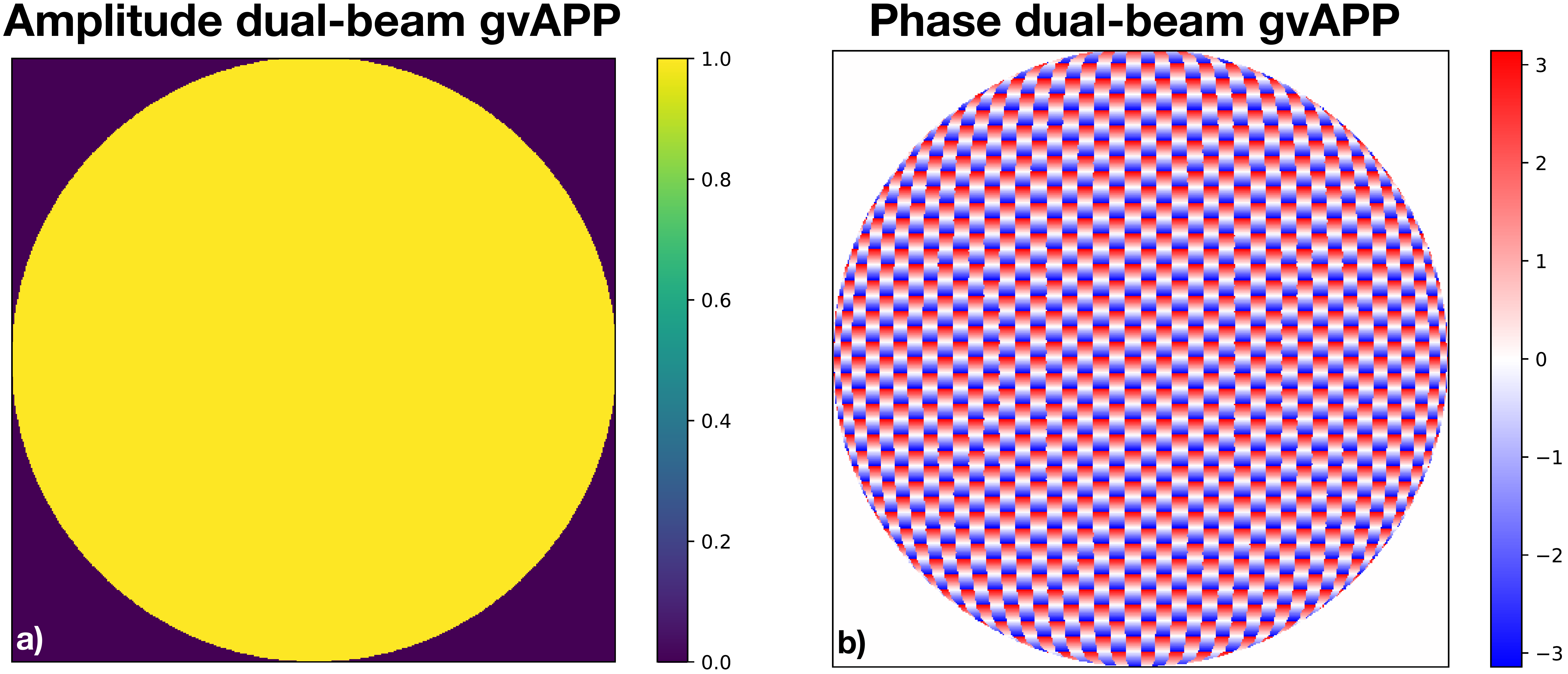}
\end{tabular}
\end{center}
\caption{Pupil-plane a) amplitude and b) phase design of the dual-beam gvAPP. 
}
\label{fig:dual-beam_gvAPPdesign}  
\end{figure}

\begin{figure}[!htb]
\begin{center}
\begin{tabular}{c}
\includegraphics[width=1\textwidth]{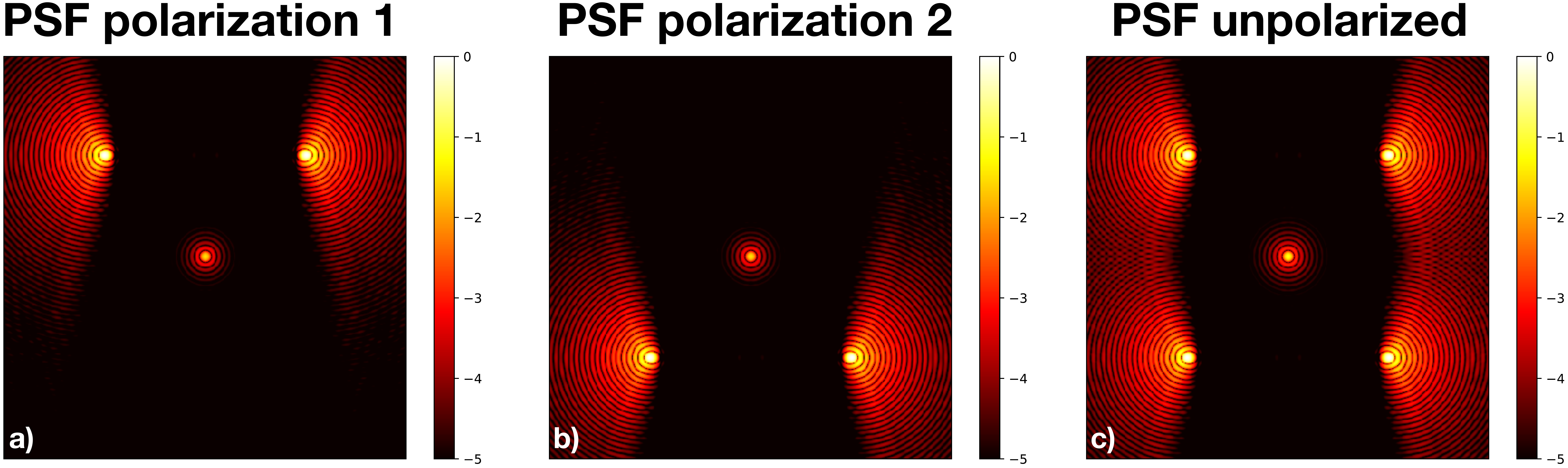}
\end{tabular}
\end{center}
\caption{
The coronagraphic PSFs for the a) first and b) second polarization state. 
c) The PSFs for unpolarized light. 
The central PSF is the leakage term. 
}
\label{fig:dual-beam_gvAPP_PSFs}  
\end{figure}

\begin{figure}[!htb]
\begin{center}
\begin{tabular}{c}
\includegraphics[width=0.75\textwidth]{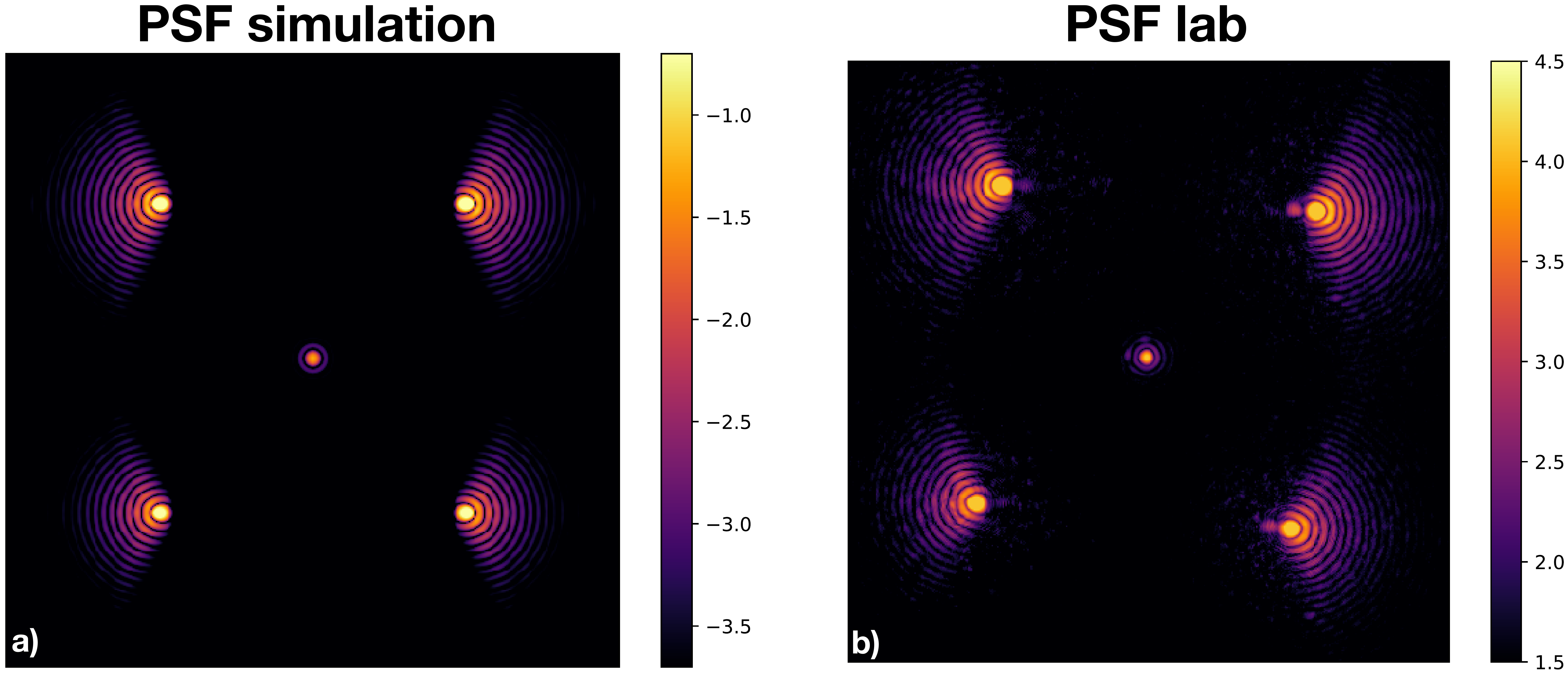}
\end{tabular}
\end{center}
\caption{
a) Simulated dual-beam gvAPP PSF with a 27\% degree of circular polarization, plotted on a scale to match the lab results.
b) Lab results of the prototype. 
}
\label{fig:dual-beam_gvAPP_data}  
\end{figure}

%
Ref~\citenum{snik2014combining} presented vAPP designs for sensitive dual-beam imaging polarimetry. 
However, these designs require a vAPP, polarization modulator, two Wollaston prims and three QWPs, making them complicated to implement into instruments, and sensitive to leakage terms. 
Here we present a simplified dual-beam polarimetric gvAPP design that sacrifices bandwidth for ease of implementation. 
It requires a gvAPP,  polarization modulator and one QWP.  
The phase and amplitude design of the gvAPP are shown in \autoref{fig:dual-beam_gvAPPdesign}. 
The phase design generates two coronagraphic PSFs with opposite dark holes per circular polarization state. 
The coronagraphic PSFs are positioned off center such that they do not overlap with the PSFs of the opposite polarization state. 
This is shown in \autoref{fig:dual-beam_gvAPP_PSFs}. 
Both polarization states now have two identical coronagraphic PSFs that can be subtracted for dual-beam polarimetry. 
We have manufactured a prototype that was tested in the lab. 
The measured PSFs and a simulation comparison is shown in \autoref{fig:dual-beam_gvAPP_data}. 
The PSFs shows show excellent agreement (barring the ghost in the lab data), which proves that the complicated phase design of \autoref{fig:dual-beam_gvAPPdesign}b can be manufactured. \\ 

Polarimetry with the dual-beam polarimetric gvAPP has the advantage that it is easily implemented with only a few components, and is relatively unaffected by leakage problems. 
The disadvantages are that the design is only suitable for observations with narrowband filters or integral-field spectrographs, and that the two coronagraphic PSFs per polarization also generate higher-order copies, which lead to light loss.
However, it might be possible to suppress the higher order copies with more advanced designs. 
\section{Unpolarized gAPvAPP}\label{sec:unpol_gvAPP}
%
As shown in \autoref{fig:ExplanationvAPPgvAPP}b gvAPPs generate two (or more) coronagraphic PSFs that are circular polarized.
This makes the relative brightness of the coronagraphic PSFs dependent on the degree of circular polarization, which can be several tens of percents for measurements with artificial sources. 
This for example complicates model-based wavefront sensing algorithms as they now also have to model and measure the degree of circular polarization, adding another error source. 
This also makes that the gvAPP is not easily integrated into a polarimeter, because you can't simple add a polarizing beamsplitter behind the system as the polarization information is already lost due to the polarization grating. 
To make polarimetry work with the gvAPP, it requires an upstream QWP and fast polarization modulator and only allows for single-beam polarimetry. 
This is extensively discussed in Ref~\citenum{bos2018fully}.  
Therefore, a gvAPP design that yields unpolarized coronagraphic PSFs would be of great benefit. \\
\begin{figure}[!htb]
\begin{center}
\begin{tabular}{c}
\includegraphics[width=0.75\textwidth]{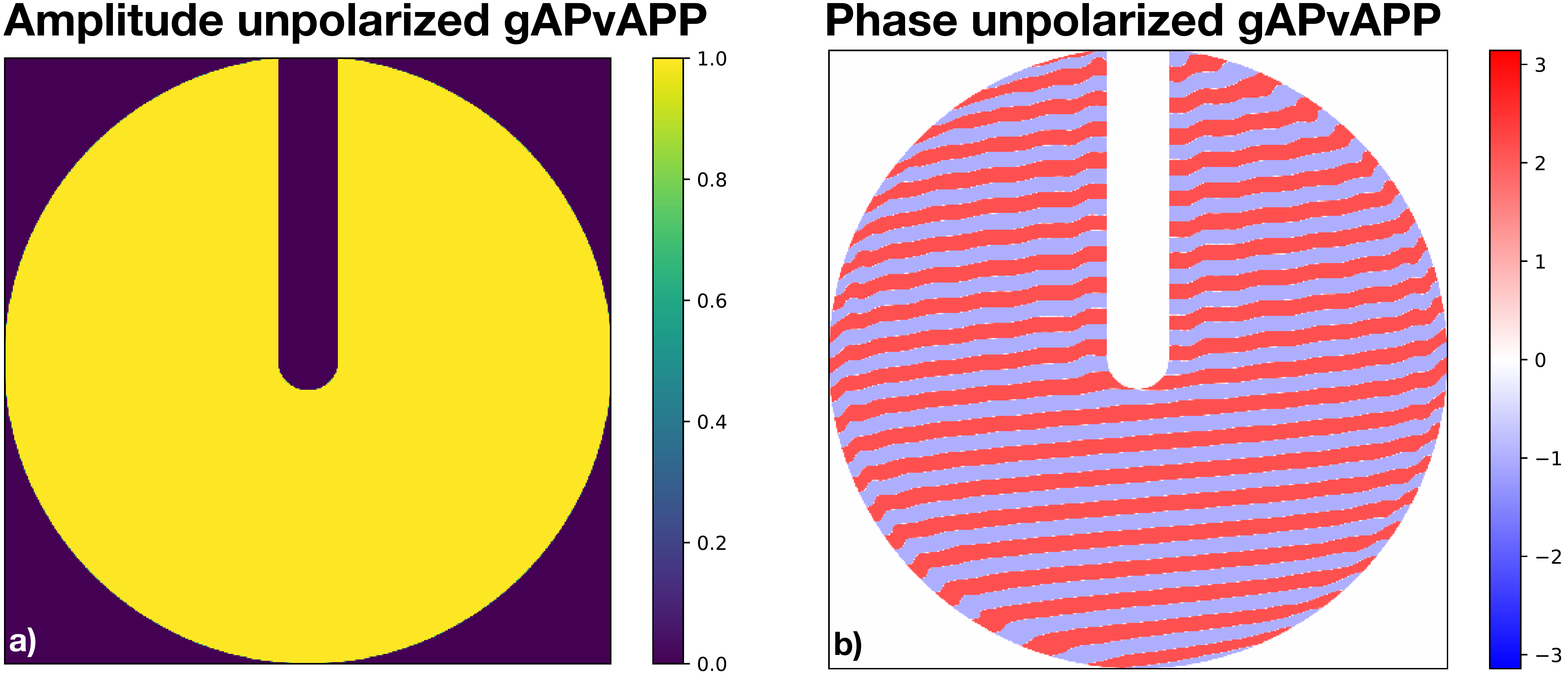}
\end{tabular}
\end{center}
\caption{Pupil-plane a) amplitude and b) phase design of the unpolarized gAPvAPP. 
}
\label{fig:unpol_gAPvAPPdesign}  
\end{figure} 
\begin{figure}[!htb]
\begin{center}
\begin{tabular}{c}
\includegraphics[width=1\textwidth]{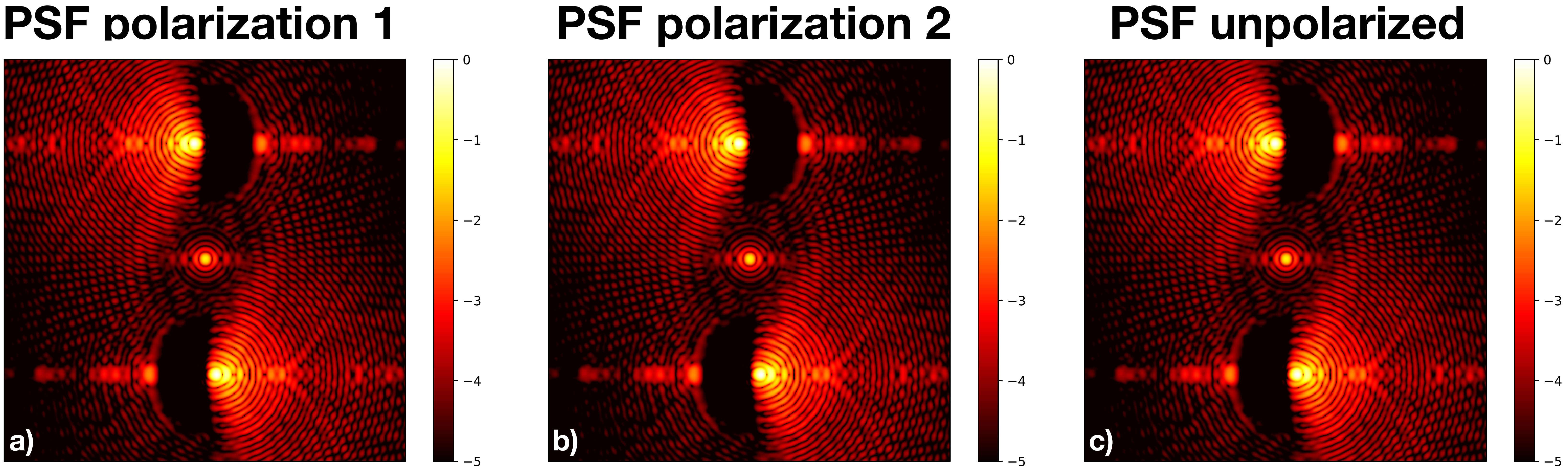}
\end{tabular}
\end{center}
\caption{
The coronagraphic PSFs for the a) first and b) second polarization state. 
c) The PSFs for unpolarized light. 
}
\label{fig:unpol_gAPvAPP_PSFs}  
\end{figure} 
\begin{figure}[!htb]
\begin{center}
\begin{tabular}{c}
\includegraphics[width=0.75\textwidth]{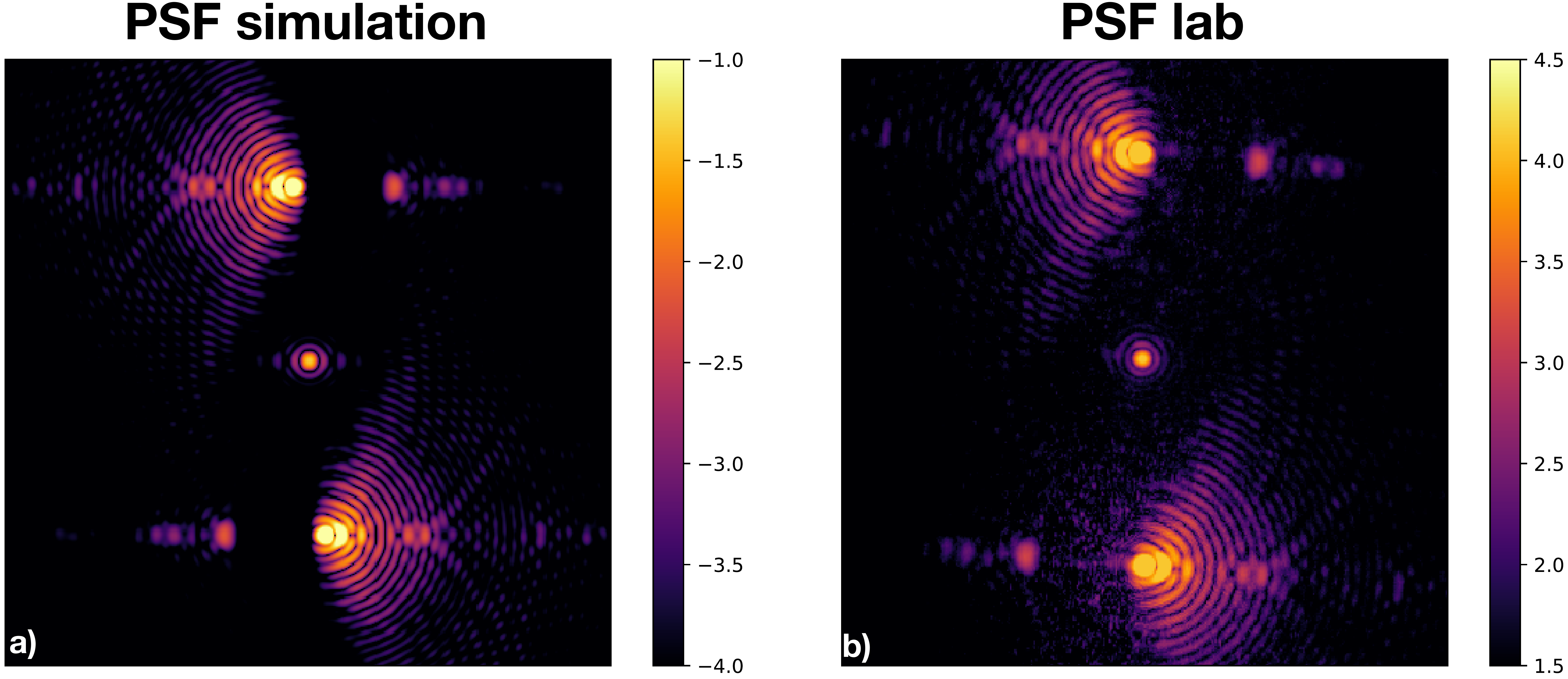}
\end{tabular}
\end{center}
\caption{
a) Simulated dual-beam unpolarized gAPvAPP PSF, plotted on a scale to match the lab results.
b) Lab results of the prototype. 
}
\label{fig:unpol_gAPvAPP_data}  
\end{figure} 

Here we present an unpolarized gAPvAPP design, which is plotted in \autoref{fig:unpol_gAPvAPPdesign}.
This design generates for both circular polarization states two, off-axis coronagraphic PSFs. 
The entire system of the two coronagraphic PSFs is point symmetric, which is necessary to make sure that the opposite circular polarization state precisely overlaps with the same, but flipped PSFs.
Otherwise, the degree of circular polarization would not be zero. 
Simulated PSFs of the presented design are shown in \autoref{fig:unpol_gAPvAPP_PSFs}. 
The simulations show that the PSFs for the two circular polarization states and for unpolarized light look the same. 
We also manufactured a prototype that was tested in the lab. 
The resulting PSF and a numerical PSF as comparison are shown in \autoref{fig:unpol_gAPvAPP_data}.
The simulation and the lab PSF are qualitatively very similar, which shows that these designs can be manufactured. 
The measured normalized brightness differences (or degree of circular polarization) between the two PSFs for the unpolarized gvAPvAPP is on the order of $\sim$2.35\%. 
The normalized brightness differences for other gvAPPs measured in the same optical systems were $\sim$27\%, due to circular polarization effects. 
 These results show that the unpolarized gAPvAPP greatly mitigated degree of circular polarization problems. 
 The remaining $\sim$2.35\% of brightness difference could be due to differential aberrations between the two coronagraphic PSFs. \\  
 
 The unpolarized gvAPvAPP has the advantage of providing coronagraphic PSFs whose relative intensity is unaffected by the degree of circular polarization. 
 This is of great benefit for focal-plane wavefront sensing and polarimetry. 
 However, this comes at the cost of higher order copies of the PSFs that are generated as well, leaking away light.
 These effect could be minimized with more advanced designs. 
 Another disadvantage is that, due to the polarization grating, these can only operate in narrowband filters or integral-field spectrographs. 
\section{Multiplexed APvAPP}\label{sec:multiplexed_APvAPP}
%
Current APvAPPs are designed to measure pupil-plane phase aberrations as these are the dominant source for speckle noise, which is one of the current limitations for HCI. 
However, at some level pupil-plane amplitude aberrations also become relevant. 
For example, amplitude aberrations play an important role in the formation of the ``asymmetric wind driven halo"\cite{cantalloube2018origin}. 
Therefore, it is important that these aberrations can be measured as well. 
Current gvAPvAPP designs do not have the required diversity in the coronagraphic PSFs to measure pupil-plane phase and amplitude simultaneously. 
Additional diversity could be provided by taking another defocussed image and include that in the wavefront measurement process.
However, this lowers the science duty cycle, and thus it is of great interest to come up with gAPvAPP designs that can measure amplitude and phase in one in-focus image. \\ 
\begin{figure}[!htb]
\begin{center}
\begin{tabular}{c}
\includegraphics[width=0.75\textwidth]{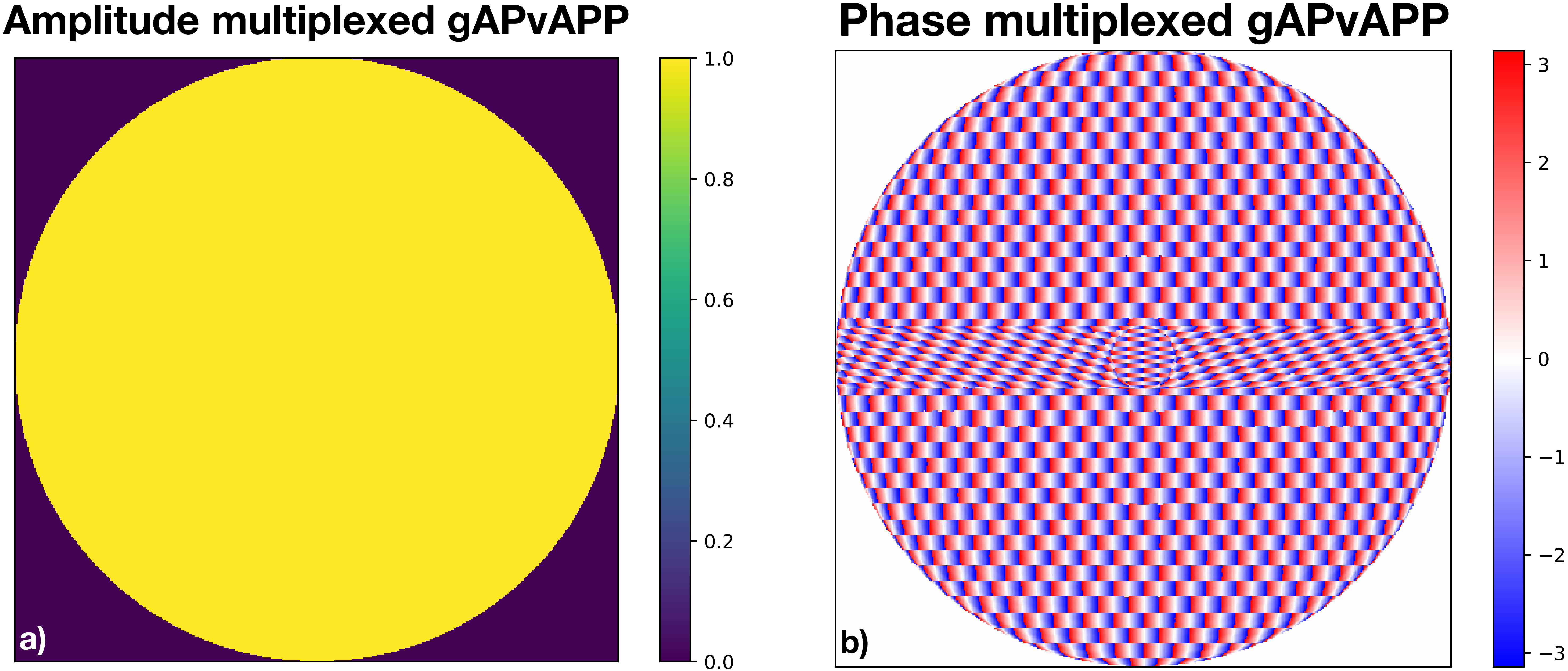}
\end{tabular}
\end{center}
\caption{Pupil-plane a) amplitude and b) phase design of the multiplexed gAPvAPP. 
}
\label{fig:multiplexed_gAPvAPPdesign}  
\end{figure} 
\begin{figure}[!htb]
\begin{center}
\begin{tabular}{c}
\includegraphics[width=1\textwidth]{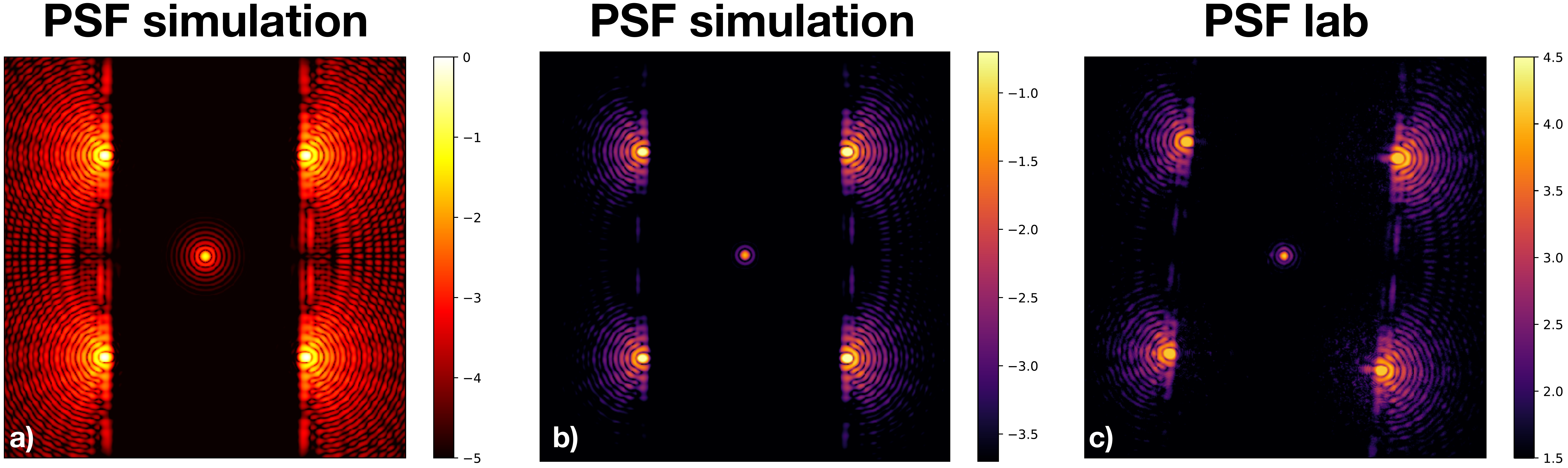}
\end{tabular}
\end{center}
\caption{
a) Simulated multiplexed gAPvAPP PSF with high dynamic range and no polarization effects.
b) Simulated multiplexed gAPvAPP PSF with a 27\% degree of circular polarization, plotted on a scale to match the lab results.
c) Lab PSF of the multiplexed gAPvAPP.
}
\label{fig:multiplexed_gAPvAPP_PSFs}  
\end{figure} 
Here we present a new gAPvAPP design that generates two coronagraphic PSFs per polarization, each with the bright field covering the same spatial frequencies, but differently probing the focal-plane electric field.
The design in presented in \autoref{fig:multiplexed_gAPvAPPdesign}, and is referred to as the multiplexed gAPvAPP.
It generates two coronagraphic PSFs that are designed with different pupil-plane amplitude asymmetries. 
These two PSFs now have enough diversity to probe phase and amplitude aberrations simultaneously, and the PSFs for the opposite polarization state cover the opposite spatial frequencies.  
Simulations and lab PSFs are shown in \autoref{fig:multiplexed_gAPvAPP_PSFs}.
It shows an excellent qualitative agreement between simulation and lab. \\  

The multiplexed gAPvAPP offers the advantage of measurement both pupil-plane phase and amplitude with one image, making it a highly efficient wavefront sensor. 
However, as with other gvAPP designs that generate multiple PSFs per polarization, the multiplexed gAPvAPP losses light due to higher order copies.
Again, we hope to mitigate this with more advanced designs. 
Also for this gvAPP, it only operates in narrowband filters and integral-field spectrographs.

\section{gSAMvAPP}\label{sec:gSAMvAPP}
%
The focal-plane wavefront sensing capabilities of gAPvAPPs come from pupil-plane amplitude asymmetries. 
If the asymmetry is not a not a natural part of the pupil, it throws away a (small) part of the light. 
While enabling focal-plane wavefront sensing is a great benefit, the asymmetry still lowers the planet throughput, affecting planetary yield calculations. 
Therefore, new gAPvAPP designs that utilizes the light from the asymmetry in some way are preferred. \\
\begin{figure}[!htb]
\begin{center}
\begin{tabular}{c}
\includegraphics[width=0.75\textwidth]{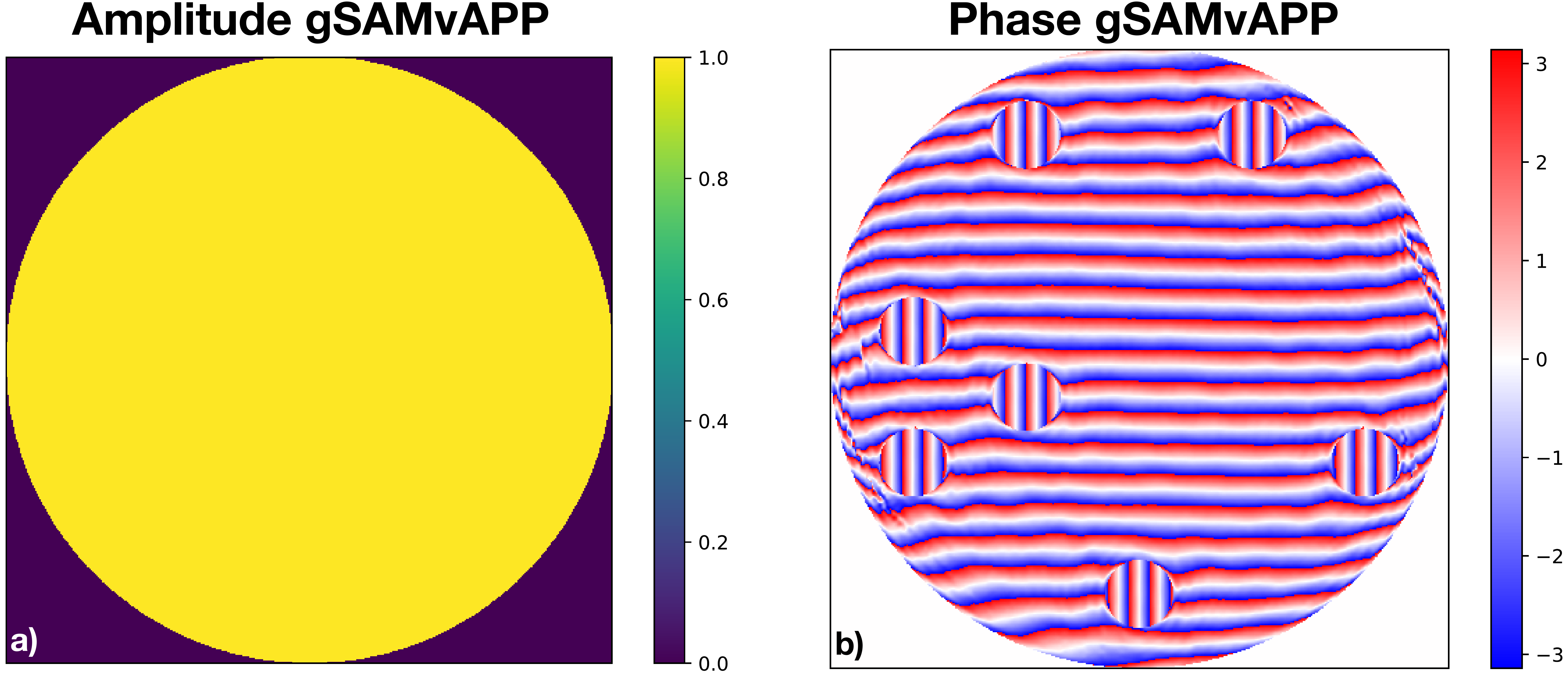}
\end{tabular}
\end{center}
\caption{Pupil-plane a) amplitude and b) phase design of the gSAMvAPP. 
}
\label{fig:gSAMvAPPdesign}  
\end{figure} 

\begin{figure}[!htb]
\begin{center}
\begin{tabular}{c}
\includegraphics[width=1\textwidth]{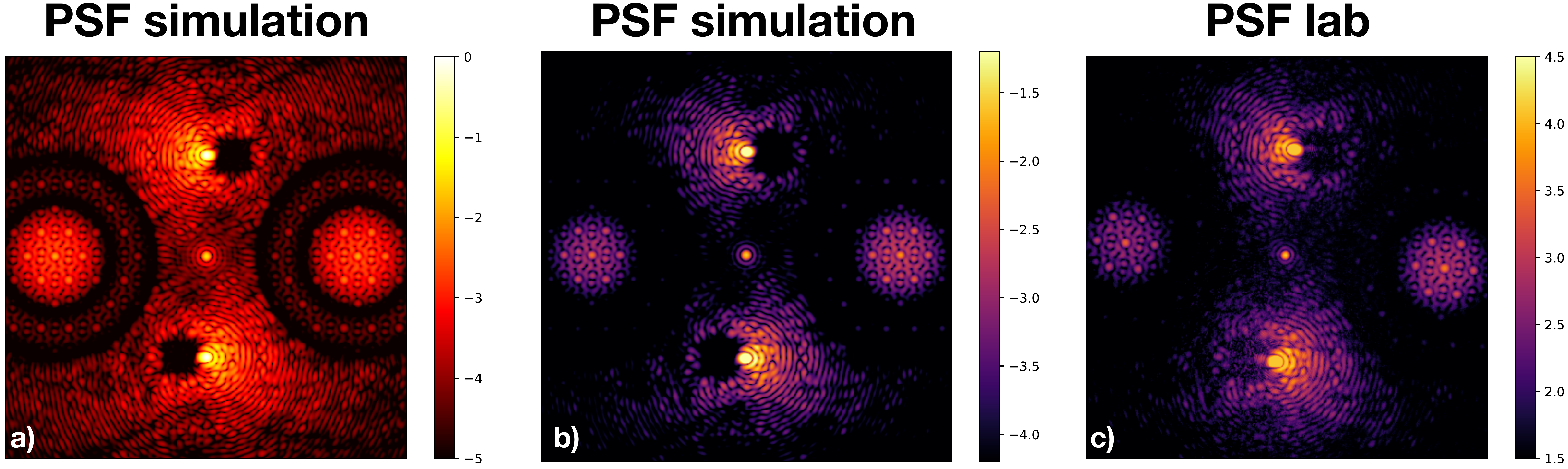}
\end{tabular}
\end{center}
\caption{
a) Simulated gSAMvAPP PSF with high dynamic range and no polarization effects.
b) Simulated gSAMvAPP PSF with a 27\% degree of circular polarization, plotted on a scale to match the lab results.
c) Lab PSF of the gSAMvAPP.
}
\label{fig:gSAMvAPP_PSFs}  
\end{figure} 

Here we present a combination of sparse aperture mask (SAM) and gvAPP, and is referred to as the gSAMvAPP. 
The design is shown in \autoref{fig:gSAMvAPPdesign}. 
A SAM is cut out from the pupil-plane amplitude with a polarization grating, following the SAM design for VLT/SPHERE\cite{cheetham2016sparse}. 
The pupil that is left over is highly asymmetric and can therefore be used for focal-plane wavefront sensing. 
The simulated coronagraphic PSF is shown in \autoref{fig:gSAMvAPP_PSFs}a. 
It shows two coronagraphic PSFs and two SAM PSFs. 
We also manufactured a prototype and tested it in the lab. 
The results are shown in \autoref{fig:gSAMvAPP_PSFs}b,c. 
The simulated and lab PSF show excellent qualitative agreement. \\ 

The gSAMvAPP combines coronagraphy with focal-plane wavefront sensing and SAM interferometry. 
This allows for a simultaneous the search for exoplanets on scales reachable by coronagraphs and interferometers, while maintaining a stable wavefront.  
The current SAM design was not optimized in any way for the combination with coronagraphy and wavefront sensing, and future design should look into holographic aperture masking approaches\cite{doelman2018multiplexed}.  
The disadvantages are that the coronagraphic performance (planet throughput) is currently strongly affected by the SAM design, and should be further optimized. 
As with all gvAPP design with multiple PSF, light is lost into higher order copies, and due to the polarization grating the designs can only be used for narrowband filters or integral-field spectrographs. 
%
\section{Conclusion}\label{sec:conclusion}
In this work we have presented six new concepts in vector-Apodizing Phase Plate coronagraphy. 
These designs deal with low-leakage solutions for broadband coronagraphy, and the integration of coronagraphy with focal-plane wavefront sensing, (dual-beam) polarimetry and sparse aperture mask interferometry. 
We have shown that creative combinations of holography and geometric phase can replace complicated polarimetric systems.  
For most of the new concepts we manufactured prototypes that were tested in the lab. 
These all provided qualitatively satisfactory PSFs, which shows that the rather complex phase design can be manufactured with high fidelity. 
With these new concepts in vAPP coronagraphy, we provide new input for system wide integration of all submodules in the HCI instrument to improve the on-sky performance. 

\acknowledgments 
We thank Alex Tripsas for his help with building the lab setup. 
The research of Steven P. Bos, David S. Doelman, and Frans Snik leading to these results has received funding from the European Research Council under ERC Starting Grant agreement 678194 (FALCONER). This research made use of HCIPy, an open-source object-oriented framework written in Python for performing end-to-end simulations of high-contrast imaging instruments \cite{por2018hcipy}.

\bibliography{report} 
\bibliographystyle{spiebib} 

\end{document}